\documentclass[sigconf]{acmart}

\AtBeginDocument{%
  }

\setcopyright{acmlicensed}
\copyrightyear{2024}
\acmYear{2024}
\acmDOI{XXXXXXX.XXXXXXX}

\acmConference[MuC '24]{Envisioning the Future of Accessible Immersive Technology}{September 1st, 2024}{Karlsruhe, Germany}
\acmBooktitle{Mensch und Computer 2024 Workshop: Envisioning the Future of Accessible Immersive Technology,
 September 1st, 2024, Karlsruhe, Germany}
\begin{document}

\title{Challenges in Mixed Reality in Assisting Adults with ADHD Symptoms}

\author{Valerie Tan}
\email{valerie.tan@tu-dortmund.de}
\orcid{0009-0005-8159-4027}
\affiliation{%
  \institution{TU Dortmund University}
  \city{Dortmund}
  \country{Germany}
}

\author{Jens Gerken}
\email{jens.gerken@tu-dortmund.de}
\orcid{https://orcid.org/0000-0002-0634-3931}
\affiliation{%
  \institution{TU Dortmund University}
  \city{Dortmund}
  \country{Germany}
}


\begin{abstract}
In this position paper, we discuss symptoms of attention deficit hyperactivity disorder (ADHD) in adults, as well as available forms of treatment or assistance in the context of mixed reality. Mixed reality offers many potentials for assisting adults with symptoms commonly found in (but not limited to) ADHD, but the availability of mixed reality solutions is not only limited commercially, but also limited in terms of proof-of-concept prototypes. We discuss two major challenges with attention assistance using mixed reality solutions: the limited availability of adult-specific prototypes and studies, as well as the limited number of solutions that offer continuous intervention of ADHD-like symptoms that users can employ in their daily life. 
\end{abstract}




\received{30 June 2024}

\maketitle

\section{Introduction}
An estimated 2-4\% of the adult population in the world are believed to have attention deficit hyperactivity disorder (ADHD)---an arguably debated yet also under-diagnosed condition \cite{Weibel2020}. ADHD is characterized by inattention, hyperactivity, and impulsivity \cite{Weibel2020, Sjwall2012}. Furthermore, such symptoms associated with ADHD are not only prevalent amongst the population with diagnosed ADHD, but amongst people with other conditions---including but not limited to---autism, anxiety disorders, depressive disorders, mood disorders, and obsessive compulsive disorder\cite{Wang2022, Weibel2020}.  

Treatment and management of ADHD typically involves pharmacological treatment and psychosocial interventions \cite{Cuber2024, Weibel2020, Corrigan2023}. Psychosocial interventions consist of three groupings of therapies according to Corrigan et al.'s systemic review on virtual reality treatment for children with ADHD \cite{Corrigan2023}. The first grouping focuses on therapies relating to contingency management in which positive reinforcement is used for positive behavioral change \cite{Petry2011, Corrigan2023}. In the context of children with ADHD, this group includes behavioral parent training, behavioral classroom interventions, and behavioral peer interventions \cite{Corrigan2023}. 
%cognitive remediation (behavioral interventions) 
The second grouping consists of cognitive behavioral therapy (CBT) as well as training interventions \cite{Corrigan2023}
Cognitive Behavioral Therapy (CBT) is one of the most common forms of cognitive remediation for ADHD, but is resource-intensive, as it requires many sessions with a specialty therapist over an extended period of time \cite{Goharinejad2022, Weibel2020}. 
Finally, the third grouping entails mindfulness and dialectical behavioral therapy intended to improve emotional regulation and reduce dysfunctional behaviors \cite{Corrigan2023, chapman2006dialectical}. 
%which therapies are proven to be most effective 
From this list of therapies, Corrigan et al. \cite{Corrigan2023} found that behavioral parent training, behavioral classroom interventions, behavioral peer interventions, and organizational skills training bring empirically-proven ``moderate improvement in ADHD symptoms'' in children. 

In this position paper, we mainly focus on the symptoms that are typically associated with ADHD in adults such as executive dysfunction and inattention, and include \textit{all} adults whose needs involve assistance in these aspects, even if they do not have a specific diagnosis or condition. 
%could add the reasoning behind this (underdiagnosis, lack of knowledge, difficulty in diagnosis and care, lack of access to care, etc)
%\section{Attention Deficit Hyperactivity Disorder and Related Conditions}

\section{Mixed Reality for ADHD}
There are a variety of approaches for using mixed reality devices for ADHD treatment. One prominent use is for diagnosing ADHD, typically through a VR-based continuous performance test (CPT) \cite{Goharinejad2022}. A second general use for mixed reality is in cognitive training through the means of serious games \cite{Ou2020} or through specific self-contained applications or simulations that train working memory, eye contact, and attention \cite{Goharinejad2022}. Other approaches include simulating a VR environment \cite{Cuber2024, Kwan2022}, or adding social presence and support via an animal-like robot \cite{OConnell2024}. Mixed Reality development overall is still new and expensive, and therefore there are a limited number of known of mixed reality treatments for ADHD-like symptoms, even if we count prototypes and proof of concepts \cite{Goharinejad2022}.

% why even use mr for this 
Such mixed reality treatments for ADHD have potential benefits. For instance, gamification may evoke enjoyment and therefore increase intrinsic motivation in users \cite{Goharinejad2022, Cuber2024}. Mixed reality treatments also offer a potential solution to the low availability of time-intensive traditional in-person therapy methods, as well as enable patients to work in a comfortable environment \cite{Cuber2024, Goharinejad2022}.

\section{Challenges and Further Research}
\subsection{Attention Assistance for Adults}
Psychiatric and psychological treatment for adults with ADHD has been long been extremely limited, as ADHD has been incorrectly perceived as a childhood condition \cite{Weibel2020, ginsberg2014underdiagnosis}. Adults with ADHD, who were diagnosed as children, often do not continue treatment into their adulthood \cite{ginsberg2014underdiagnosis}. Meanwhile, many adults with ADHD remain undiagnosed and therefore untreated \cite{ginsberg2014underdiagnosis}. Given the gap in the general treatment of adults with ADHD-symptoms, it is unsurprising to us that at time the writing of this position paper, we have currently found a very limited number of mixed reality prototypes geared specifically towards adults. 
Additionally, Cuber et al. \cite{Cuber2024}, found that the use of virtual reality for ADHD is mostly geared towards the assessment or diagnosis of ADHD, as opposed to support. 
%We found that most mixed reality prototypes and studies related to ADHD and related conditions have been geared towards children and/or the assessment, but not treatment of ADHD \cite{Cuber2024}. 
While both of these aspects related to mixed reality for ADHD are nonetheless very important, we find that further research on developing support for adults with ADHD symptoms is necessary. 
ADHD and related conditions overall manifest differently in adults. For example, adults with ADHD tend to have less hyperactivity in adulthood, but maintain struggles with inattention \cite{Weibel2020}. Cuber et al. \cite{Cuber2024} note that existing works focus more on having participants perform ``narrowly defined'' tasks intended to train certain skills, as opposed to assisting adults with real-world tasks. We also agree with their assessment, and argue that further research should include solutions that allow people to integrate treatment or generalized assistance into their everyday life. 
% cost and ofc development in general 
\subsection{Continuous Intervention}
Most mixed reality treatment solutions that we found involve a task or session that mimic a therapy session. This form of treatment is related to internet-administered CBT, and older form which Linder \cite{Lindner2020} describes as ``in essence a digital form of bibliotherapy: a virtual self-help book where modules replace chapters, delivered not on paper but via an online platform, with or without support from an online therapist (often through asynchronous messaging).'' While this form of treatment is essential and should be developed further in the context of mixed reality, we should also consider intervention and assistance in daily life. Behavioral interventions for children with ADHD, including behavioral parent training, behavioral peer interventions, and behavioral classroom interventions have been proven to be effective \cite{Corrigan2023}. Prototyping and testing adult versions of these forms of therapy may therefore be beneficial. These behavioral treatments for children are already resource-intensive, as they require long-term monitoring from parents, teachers, and peers; however, options for independent adults are inherently limited  \cite{Cuber2024, Corrigan2023}. As far as we are aware, there are very few available prototypes or studies on solutions that continuously monitor or assist adults with ADHD-like symptoms, detect inattention, or provide feedback. As mentioned before, Cuber et al.'s VR environment setup includes a feedback bar that displays the system's estimated user input level \cite{Cuber2024}. O'Connell et al. \cite{OConnell2024} use a social assistive robot with continuous movements coupled with a touchscreen to provide a social presence companion for college-aged students as they study or work on assignments. Further research could similarly make use of the unique aspects of mixed reality. For instance, eye tracking technologies could be used for monitoring and detection \cite{Wang2022} of user attention. We could explore this idea further in future augmented reality devices that are capable of integrating the real-world space with visual and virtual forms of interaction and assistance. In addition, wearables have the potential of gathering biometric data, which could be integrated into solutions. 
%\subsection{Longitudinal Studies}
\section{Conclusion}
Current mixed reality prototypes intended to treat or assist adults with ADHD-like symptoms are limited in terms of demographics (treating the condition as a mainly childhood condition but not one found in adulthood) as well the availability of solutions that can be integrated into daily life with real-world activities. In addition, the sheer cost and novelty of mixed reality is currently another barrier. We can however interpret these challenges and limitations optimistically as possibilities for future solutions that may empower all adults with symptoms related to ADHD.

\bibliographystyle{ACM-Reference-Format}
\bibliography{citations}


\begin{thebibliography}{13}

%%% the Web address.

\ifx \showCODEN    \undefined \def \showCODEN     #1{\unskip}     \fi
\ifx \showDOI      \undefined \def \showDOI       #1{#1}\fi
\ifx \showISBNx    \undefined \def \showISBNx     #1{\unskip}     \fi
\ifx \showISBNxiii \undefined \def \showISBNxiii  #1{\unskip}     \fi
\ifx \showISSN     \undefined \def \showISSN      #1{\unskip}     \fi
\ifx \showLCCN     \undefined \def \showLCCN      #1{\unskip}     \fi
\ifx \shownote     \undefined \def \shownote      #1{#1}          \fi
\ifx \showarticletitle \undefined \def \showarticletitle #1{#1}   \fi
\ifx \showURL      \undefined \def \showURL       {\relax}        \fi
% invisible to TeX
\providecommand\bibfield[2]{#2}
\providecommand\bibinfo[2]{#2}
\providecommand\natexlab[1]{#1}
\providecommand\showeprint[2][]{arXiv:#2}

\bibitem[Chapman(2006)]%
        {chapman2006dialectical}
\bibfield{author}{\bibinfo{person}{Alexander~L Chapman}.} \bibinfo{year}{2006}\natexlab{}.
\newblock \showarticletitle{Dialectical behavior therapy: Current indications and unique elements}.
\newblock \bibinfo{journal}{\emph{Psychiatry (Edgmont)}} \bibinfo{volume}{3}, \bibinfo{number}{9} (\bibinfo{year}{2006}), \bibinfo{pages}{62}.
\newblock


\bibitem[Corrigan et~al\mbox{.}(2023)]%
        {Corrigan2023}
\bibfield{author}{\bibinfo{person}{Niamh Corrigan}, \bibinfo{person}{Costina-Ruxandra Păsărelu}, {and} \bibinfo{person}{Alexandra Voinescu}.} \bibinfo{year}{2023}\natexlab{}.
\newblock \showarticletitle{Immersive virtual reality for improving cognitive deficits in children with ADHD: a systematic review and meta-analysis}.
\newblock \bibinfo{journal}{\emph{Virtual Reality}} \bibinfo{volume}{27}, \bibinfo{number}{4} (\bibinfo{date}{Feb.} \bibinfo{year}{2023}), \bibinfo{pages}{3545–3564}.
\newblock
\showISSN{1434-9957}
\urldef\tempurl%
\url{https://doi.org/10.1007/s10055-023-00768-1}
\showDOI{\tempurl}


\bibitem[Cuber et~al\mbox{.}(2024)]%
        {Cuber2024}
\bibfield{author}{\bibinfo{person}{Isabelle Cuber}, \bibinfo{person}{Juliana~G Goncalves De~Souza}, \bibinfo{person}{Irene Jacobs}, \bibinfo{person}{Caroline Lowman}, \bibinfo{person}{David Shepherd}, \bibinfo{person}{Thomas Fritz}, {and} \bibinfo{person}{Joshua~M Langberg}.} \bibinfo{year}{2024}\natexlab{}.
\newblock \showarticletitle{Examining the Use of VR as a Study Aid for University Students with ADHD}. In \bibinfo{booktitle}{\emph{Proceedings of the CHI Conference on Human Factors in Computing Systems}} (Honolulu, HI, USA) \emph{(\bibinfo{series}{CHI '24})}. \bibinfo{publisher}{Association for Computing Machinery}, \bibinfo{address}{New York, NY, USA}, Article \bibinfo{articleno}{65}, \bibinfo{numpages}{16}~pages.
\newblock
\showISBNx{9798400703300}
\urldef\tempurl%
\url{https://doi.org/10.1145/3613904.3643021}
\showDOI{\tempurl}


\bibitem[Ginsberg et~al\mbox{.}(2014)]%
        {ginsberg2014underdiagnosis}
\bibfield{author}{\bibinfo{person}{Ylva Ginsberg}, \bibinfo{person}{Javier Quintero}, \bibinfo{person}{Ernie Anand}, \bibinfo{person}{Marta Casillas}, {and} \bibinfo{person}{Himanshu~P Upadhyaya}.} \bibinfo{year}{2014}\natexlab{}.
\newblock \showarticletitle{Underdiagnosis of attention-deficit/hyperactivity disorder in adult patients: a review of the literature}.
\newblock \bibinfo{journal}{\emph{The primary care companion for CNS disorders}} \bibinfo{volume}{16}, \bibinfo{number}{3} (\bibinfo{year}{2014}), \bibinfo{pages}{23591}.
\newblock


\bibitem[Goharinejad et~al\mbox{.}(2022)]%
        {Goharinejad2022}
\bibfield{author}{\bibinfo{person}{Saeideh Goharinejad}, \bibinfo{person}{Samira Goharinejad}, \bibinfo{person}{Sadrieh Hajesmaeel-Gohari}, {and} \bibinfo{person}{Kambiz Bahaadinbeigy}.} \bibinfo{year}{2022}\natexlab{}.
\newblock \showarticletitle{The usefulness of virtual, augmented, and mixed reality technologies in the diagnosis and treatment of attention deficit hyperactivity disorder in children: an overview of relevant studies}.
\newblock \bibinfo{journal}{\emph{BMC Psychiatry}} \bibinfo{volume}{22}, \bibinfo{number}{1} (\bibinfo{date}{Jan.} \bibinfo{year}{2022}).
\newblock
\showISSN{1471-244X}
\urldef\tempurl%
\url{https://doi.org/10.1186/s12888-021-03632-1}
\showDOI{\tempurl}


\bibitem[Kwan et~al\mbox{.}(2022)]%
        {Kwan2022}
\bibfield{author}{\bibinfo{person}{Ho~Yan Kwan}, \bibinfo{person}{Lang Lin}, \bibinfo{person}{Conor Fahy}, \bibinfo{person}{Jethro Shell}, \bibinfo{person}{Shiqi Pang}, {and} \bibinfo{person}{Yongkang Xing}.} \bibinfo{year}{2022}\natexlab{}.
\newblock \showarticletitle{Designing VR training systems for children with attention deficit hyperactivity disorder (ADHD)}. In \bibinfo{booktitle}{\emph{2022 IEEE Conference on Virtual Reality and 3D User Interfaces Abstracts and Workshops (VRW)}}. \bibinfo{pages}{88--89}.
\newblock
\urldef\tempurl%
\url{https://doi.org/10.1109/VRW55335.2022.00030}
\showDOI{\tempurl}


\bibitem[Lindner(2020)]%
        {Lindner2020}
\bibfield{author}{\bibinfo{person}{Philip Lindner}.} \bibinfo{year}{2020}\natexlab{}.
\newblock \showarticletitle{Better, Virtually: the Past, Present, and Future of Virtual Reality Cognitive Behavior Therapy}.
\newblock \bibinfo{journal}{\emph{International Journal of Cognitive Therapy}} \bibinfo{volume}{14}, \bibinfo{number}{1} (\bibinfo{date}{Oct.} \bibinfo{year}{2020}), \bibinfo{pages}{23–46}.
\newblock
\showISSN{1937-1217}
\urldef\tempurl%
\url{https://doi.org/10.1007/s41811-020-00090-7}
\showDOI{\tempurl}


\bibitem[O'Connell et~al\mbox{.}(2024)]%
        {OConnell2024}
\bibfield{author}{\bibinfo{person}{Amy O'Connell}, \bibinfo{person}{Ashveen Banga}, \bibinfo{person}{Jennifer Ayissi}, \bibinfo{person}{Nikki Yaminrafie}, \bibinfo{person}{Ellen Ko}, \bibinfo{person}{Andrew Le}, \bibinfo{person}{Bailey Cislowski}, {and} \bibinfo{person}{Maja Mataric}.} \bibinfo{year}{2024}\natexlab{}.
\newblock \showarticletitle{Design and Evaluation of a Socially Assistive Robot Schoolwork Companion for College Students with ADHD}. In \bibinfo{booktitle}{\emph{Proceedings of the 2024 ACM/IEEE International Conference on Human-Robot Interaction}} (Boulder, CO, USA) \emph{(\bibinfo{series}{HRI '24})}. \bibinfo{publisher}{Association for Computing Machinery}, \bibinfo{address}{New York, NY, USA}, \bibinfo{pages}{533–541}.
\newblock
\showISBNx{9798400703225}
\urldef\tempurl%
\url{https://doi.org/10.1145/3610977.3634929}
\showDOI{\tempurl}


\bibitem[Ou et~al\mbox{.}(2020)]%
        {Ou2020}
\bibfield{author}{\bibinfo{person}{Yang-Kun Ou}, \bibinfo{person}{Yu-Lin Wang}, \bibinfo{person}{Hua-Cheng Chang}, \bibinfo{person}{Shih-Yin Yen}, \bibinfo{person}{Yu-Hua Zheng}, {and} \bibinfo{person}{Bih-O. Lee}.} \bibinfo{year}{2020}\natexlab{}.
\newblock \showarticletitle{Development of virtual reality rehabilitation games for children with attention-deficit hyperactivity disorder}.
\newblock \bibinfo{journal}{\emph{Journal of Ambient Intelligence and Humanized Computing}} \bibinfo{volume}{11}, \bibinfo{number}{11} (\bibinfo{date}{April} \bibinfo{year}{2020}), \bibinfo{pages}{5713–5720}.
\newblock
\showISSN{1868-5145}
\urldef\tempurl%
\url{https://doi.org/10.1007/s12652-020-01945-9}
\showDOI{\tempurl}


\bibitem[Petry(2011)]%
        {Petry2011}
\bibfield{author}{\bibinfo{person}{Nancy~M. Petry}.} \bibinfo{year}{2011}\natexlab{}.
\newblock \showarticletitle{Contingency management: what it is and why psychiatrists should want to use it}.
\newblock \bibinfo{journal}{\emph{The Psychiatrist}} \bibinfo{volume}{35}, \bibinfo{number}{5} (\bibinfo{date}{May} \bibinfo{year}{2011}), \bibinfo{pages}{161–163}.
\newblock
\showISSN{1758-3217}
\urldef\tempurl%
\url{https://doi.org/10.1192/pb.bp.110.031831}
\showDOI{\tempurl}


\bibitem[Sj\"{o}wall et~al\mbox{.}(2012)]%
        {Sjwall2012}
\bibfield{author}{\bibinfo{person}{Douglas Sj\"{o}wall}, \bibinfo{person}{Linda Roth}, \bibinfo{person}{Sofia Lindqvist}, {and} \bibinfo{person}{Lisa~B. Thorell}.} \bibinfo{year}{2012}\natexlab{}.
\newblock \showarticletitle{Multiple deficits in ADHD: executive dysfunction, delay aversion, reaction time variability, and emotional deficits}.
\newblock \bibinfo{journal}{\emph{Journal of Child Psychology and Psychiatry}} \bibinfo{volume}{54}, \bibinfo{number}{6} (\bibinfo{date}{Oct.} \bibinfo{year}{2012}), \bibinfo{pages}{619–627}.
\newblock
\showISSN{1469-7610}
\urldef\tempurl%
\url{https://doi.org/10.1111/jcpp.12006}
\showDOI{\tempurl}


\bibitem[Wang et~al\mbox{.}(2022)]%
        {Wang2022}
\bibfield{author}{\bibinfo{person}{Katherine Wang}, \bibinfo{person}{Simon~J. Julier}, {and} \bibinfo{person}{Youngjun Cho}.} \bibinfo{year}{2022}\natexlab{}.
\newblock \showarticletitle{Attention-Based Applications in Extended Reality to Support Autistic Users: A Systematic Review}.
\newblock \bibinfo{journal}{\emph{IEEE Access}}  \bibinfo{volume}{10} (\bibinfo{year}{2022}), \bibinfo{pages}{15574--15593}.
\newblock
\urldef\tempurl%
\url{https://doi.org/10.1109/ACCESS.2022.3147726}
\showDOI{\tempurl}


\bibitem[Weibel et~al\mbox{.}(2020)]%
        {Weibel2020}
\bibfield{author}{\bibinfo{person}{S. Weibel}, \bibinfo{person}{O. Menard}, \bibinfo{person}{A. Ionita}, \bibinfo{person}{M. Boumendjel}, \bibinfo{person}{C. Cabelguen}, \bibinfo{person}{C. Kraemer}, \bibinfo{person}{J.-A. Micoulaud-Franchi}, \bibinfo{person}{S. Bioulac}, \bibinfo{person}{N. Perroud}, \bibinfo{person}{A. Sauvaget}, \bibinfo{person}{L. Carton}, \bibinfo{person}{M. Gachet}, {and} \bibinfo{person}{R. Lopez}.} \bibinfo{year}{2020}\natexlab{}.
\newblock \showarticletitle{Practical considerations for the evaluation and management of Attention Deficit Hyperactivity Disorder (ADHD) in adults}.
\newblock \bibinfo{journal}{\emph{L’Encéphale}} \bibinfo{volume}{46}, \bibinfo{number}{1} (\bibinfo{date}{Feb.} \bibinfo{year}{2020}), \bibinfo{pages}{30–40}.
\newblock
\showISSN{0013-7006}
\urldef\tempurl%
\url{https://doi.org/10.1016/j.encep.2019.06.005}
\showDOI{\tempurl}


\end{thebibliography}

\end{document}